# Data-Augmented Machine Learning for Predicting Biomass-Derived Hard Carbon Anode Performance in Sodium-Ion Batteries

Gang Chen[a], Zihan Yang[b], Peng Sun[b], Chenglong Wang[a,*], Jinliang Li[b]*, Guang Yang[a], Likun Pan[a]*

[a] *Shanghai Key Laboratory of Magnetic Resonance, School of Physics and Electronic Science, Institute of Magnetic Resonance and Molecular Imaging in Medicine, East China Normal University, Shanghai 200241, China*

[b] *Siyuan Laboratory, Guangdong Provincial Engineering Technology Research Center of Vacuum Coating Technologies and New Energy Materials, Department of Physics, College of Physics & Optoelectronic Engineering, Jinan University, Guangzhou 510632, China*

*Corresponding authors

E-mail: clwang@phy.ecnu.edu.cn (C. Wang); lijinliang@email.jnu.edu.cn (J. Li); lkpan@phy.ecnu.edu.cn (L. Pan)

**Abstract：**

Biomass-derived hard carbon has become the most promising anode material for sodium-ion batteries (SIBs) due to its high capacity and excellent cycling stability. However, the effects of synthesis parameters and structural features on hard carbon's (HC) electrochemical performance are still unclear, requiring time-consuming and resource-intensive experimental investigations. Machine learning (ML) offers a promising solution by training on large datasets to predict hard carbon performance more efficiently, saving time and resources. In this study, four ML models were used to predict the capacity and initial Coulombic efficiency (ICE) of HC. Data augmentation based on the TabPFN technique was employed to improve model robustness under limited data conditions, and the relationships between features and electrochemical performance were examined. Notably, the XGBoost model achieved an $R^2$ of 0.854 and an RMSE of 23.290 mAh $g^{-1}$ for capacity prediction, and an $R^2$ of 0.868 and an RMSE of 3.813% for ICE prediction. Shapley Additive Explanations (SHAP) and Partial Dependence Plot (PDP) analyses identified carbonization temperature (Temperature_2) as the most important factor influencing both capacity and ICE. Furthermore, we used bamboo as the precursor to synthesize four hard carbons based on the predictive approach. The electrochemical performance of these samples closely matched our predictions. By leveraging machine-learning approach, this study provides an efficient framework for accelerating the screening process of biomass-derived hard carbon candidates.

**Keywords:** Sodium-ion batteries; Machine learning; Biomass-derived hard carbon; Capacity; Initial Coulombic efficiency

# 1. Introduction

In the past few decades, lithium-ion batteries (LIBs) have received widespread applications in portable electronic devices and electric vehicles [1-4]. However, they still face challenges such as the uneven distribution of lithium resources, limited reserves, and high costs, which hinder their ability to meet the sustainability and economic requirements for large-scale energy storage systems in the future [5,6]. In contrast, sodium-ion batteries (SIBs), due to the abundant availability of sodium resources, lower costs, and superior safety, are considered an ideal technology to address the needs of large-scale energy storage systems [7-10]. Nevertheless, the large ionic radius of sodium ions presents a significant challenge, making conventional graphite anodes unsuitable for SIBs [11,12]. As a result, most researchers have focused on hard carbon anodes, owing to their rich pore structures, low voltage platforms, and excellent cycling stability [13-16]. Currently, hard carbon is typically derived from biomass. However, the practical application of biomass-derived hard carbon still faces a series of substantial challenges. The variability in precursor sources, complex pyrolysis conditions, and diverse post-treatment processes leads to highly complex and non-uniform microstructures, which severely compromise the controllability and predictability of their performance [17-20]. Furthermore, key electrochemical performance indicators, such as initial Coulombic efficiency (ICE) and capacity, are largely governed by the intricate and nonlinear interactions between material microstructure and fabrication parameters [21,22]. Unfortunately, traditional trial-and-error experimental methods are costly, time-consuming, and lack systematic rigor, making it difficult to elucidate these complex structure-performance relationships. This significantly impedes the further development and industrialization of high-performance hard carbon anode materials.

In recent years, machine learning (ML) has emerged as a powerful tool in the field of energy materials, offering exceptional data analysis capabilities and high prediction efficiency [23-26]. Leveraging data-driven modeling, ML facilitates the rapid prediction of material performance, uncovers complex nonlinear relationships between structure and properties, and effectively compensates for the limitations of traditional trial-and-error approaches [27,28]. This not only enhances research efficiency but also accelerates the discovery of underlying mechanisms. For instance, Owusu et al. employed multiple ML models to predict the capacity of citrus-derived biomass-based hard carbon, where the best-performing Gradient Boosting (GB) model achieved an $R^2$ of 0.467 and a root mean square error (RMSE) of 68.413 mAh g$^{-1}$ on the test set [29]. Zhang et al. analyzed the influence of lignin content on hard carbon structure and capacity using various ML models, with the eXtreme Gradient Boosting (XGBoost) model achieving an $R^2$ of 0.60. Their analysis suggested that higher lignin content and well-defined lignin structures enhance sodium storage capacity [30]. Ji et al. employed a Bootstrap Aggregating (Bagging) model to predict the electrochemical performance of hard carbon prepared via thermomechanical coupling, obtaining an $R^2$ of 0.800 and an RMSE of 25 mAh g$^{-1}$ for capacity prediction, and an $R^2$ of 0.710 with an RMSE of 0.04 for ICE [31]. Despite these advances, the prediction accuracy of existing studies remains limited, and most of them lack essential experimental validation, reducing their practical relevance for material optimization.

In this work, we adopted ML methods to predict the capacity and ICE of hard carbons in SIBs and examined how process parameters and structural features influence their electrochemical performance. A dataset comprising 350 valid entries from 95 high-quality sources was assembled, including process parameters, structural characteristics, and electrochemical performance metrics. To overcome the challenge of limited sample size, the Tabular Prior-Data Fitted Network (TabPFN) model was used for data augmentation, improving model robustness and generalization [32]. Four popular ML models, XGBoost, Random Forest (RF), Gradient Boosting Regression (GBR), and Light Gradient Boosting Machine (LightGBM), were evaluated for their ability to predict capacity and ICE [33-36]. The results showed that the XGBoost model outperformed the others, achieving an $R^2$ of 0.854 for capacity prediction on the test set with an RMSE of 23.290 mAh g$^{-1}$, and an $R^2$ of 0.868 with an RMSE of 3.813% for ICE, surpassing models reported in previous studies. Additionally, we used Shapley Additive exPlanations (SHAP) and Partial Dependence Plot (PDP) analyses to understand how key features influence electrochemical performance. To test the practical relevance of our ML model, we carried out experimental validation of the predicted results. The findings showed a high level of agreement between the predictions, feature importance analysis, and experimental outcomes, further confirming the model's effectiveness in predicting hard carbon performance. We believe our prediction approach offers significant potential to guide the experimental optimization of hard carbon anodes in SIBs.

## 2. Results and discussion

### 2.1 Exploratory data analysis

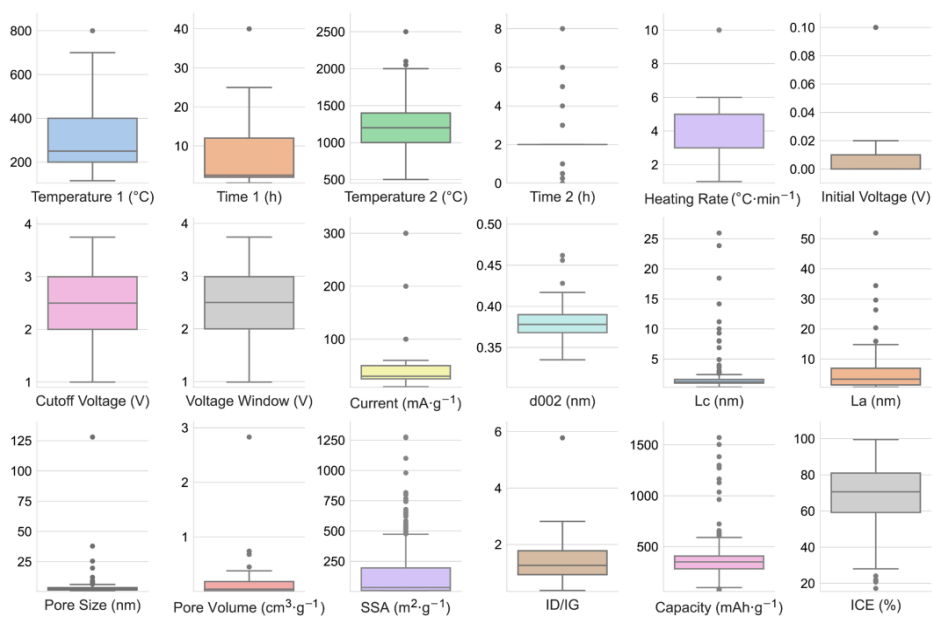

**Fig. 1** Boxplots of input features and target variables

To acquire an initial understanding of the raw dataset before preprocessing and modeling, we performed exploratory data analysis. Fig. 1 shows boxplots of all input features and target variables on their original scale (Detailed feature variables and their descriptions are listed in Table S1). Several features, such as SSA, Pore Volume, Capacity, and ICE, display pronounced right-skewed distributions with long upper tails and many outliers, indicating significant variability in material structure and electrochemical performance across studies. These long-tailed patterns suggest that some samples have exceptionally high values, potentially related to specific precursors or synthesis protocols. In contrast, structural features like d002, Lc, and ID/IG exhibit tighter, more symmetric distributions, reflecting consistent reporting across literature. Moderate variability is seen in processing-related parameters like Temperature_1 and Time_1, indicating diverse heat treatment strategies.

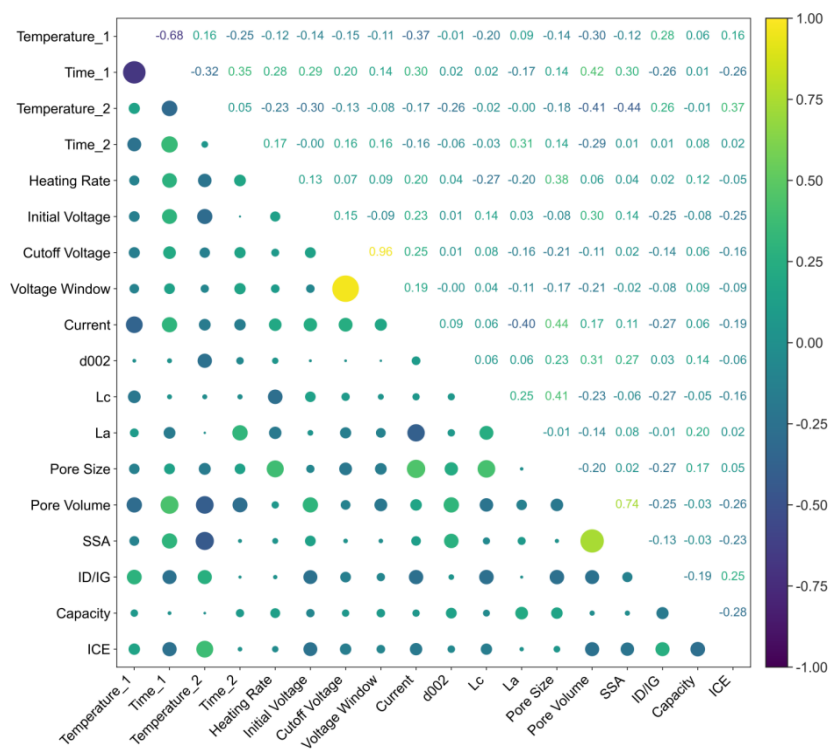

**Fig. 2** Spearman correlation matrix for all input features and target variables

We also provide the Spearman correlation matrix, which illustrates monotonic associations between features and targets, as shown in Fig. 2. A strong negative correlation (ρ=–0.68) exists between Temperature_1 and Time_1, likely indicating an inverse relationship between temperature and dwell time. SSA and Pore Volume exhibit a strong positive correlation (ρ=0.74), consistent with their shared relevance to porosity. The strongest correlation observed (ρ=0.96) is between Cutoff Voltage and Voltage Window, which are mathematically linked but were retained in subsequent modeling due to their different physical interpretations. Additionally, the ID/IG ratio shows weak correlations with most other features, suggesting it is relatively independent and potentially unique role in characterizing microstructural disorder. Notably, capacity does not show strong correlations with most features, implying that its variation is probably driven by complex, nonlinear interactions rather than direct linear relationships with individual variables. Conversely, ICE displays moderate correlations with Current, Initial Voltage, and Time_1, indicating its partial dependence on electrochemical testing conditions and thermal processing parameters.

**2.2 Model performance evaluation**

To evaluate the predictive performance of the chosen models, we developed regression models targeting the capacity and ICE of hard carbon anodes. Model training employed a 7:3 random train–test split, and hyperparameters were optimized via GridSearchCV with 5-fold cross-validation (detailed hyperparameters are provided in Supporting Information Table S2). Four ML algorithms were compared: XGBoost, RF, LightGBM, and GBR. Model performance on the test-set was assessed using five standard regression metrics: MAE, RMSE, MAPE, $R^2$, and EV (detailed information is provided in Supporting Information Section 1.2). Fig. 3 displays the test set performance of each model using donut bar charts, with Fig. 3a showing capacity prediction and Fig. 3b illustrating ICE prediction. To allow direct comparison across metrics with different scales, the error-based measures (MAE, RMSE, MAPE) were reverse-normalized to the range [0, 1] (denoted nMAE, nRMSE, and nMAPE), while $R^2$ and EV were left in their original forms. With this normalization, higher values uniformly indicate better model performance, enabling direct visual comparison across all indicators.

The performance differences among models across the two prediction tasks are further supported by the visualizations in Fig. 3 and Fig. S1–S2, as well as the numerical comparisons in Tables S3 and S4. XGBoost demonstrates the best performance in both capacity and ICE prediction, achieving the lowest MAE and RMSE values, along with the highest $R^2$ and EV scores, indicating excellent predictive accuracy and generalization. As shown in Fig. S1a and S2a, the fitted curves closely follow the ideal diagonal line (y = x), with residuals that are tightly clustered, indicating consistent reliability across the entire range of values. RF follows XGBoost, showing relatively low errors and stable fits, with notably strong performance in capacity prediction. In contrast, GBR and LightGBM exhibit larger errors, as evidenced by noticeable deviations and dispersed residuals in the plots, pointing to possible underfitting in certain regions. Nevertheless, across both accuracy- and stability-related evaluations, ICE prediction consistently outperforms capacity prediction. Specifically, the fitted curves for ICE are more tightly clustered, and the associated parameters exhibit superior performance, suggesting that ICE possesses stronger correlations and clearer underlying patterns. This trend is further corroborated in the annular bar plots presented in Fig. 3. It is found that the distribution of ICE-related scores across different models is relatively compact, whereas the normalized MAE and MAPE in capacity prediction fluctuate markedly, indicating a higher degree of uncertainty in the capacity prediction models.

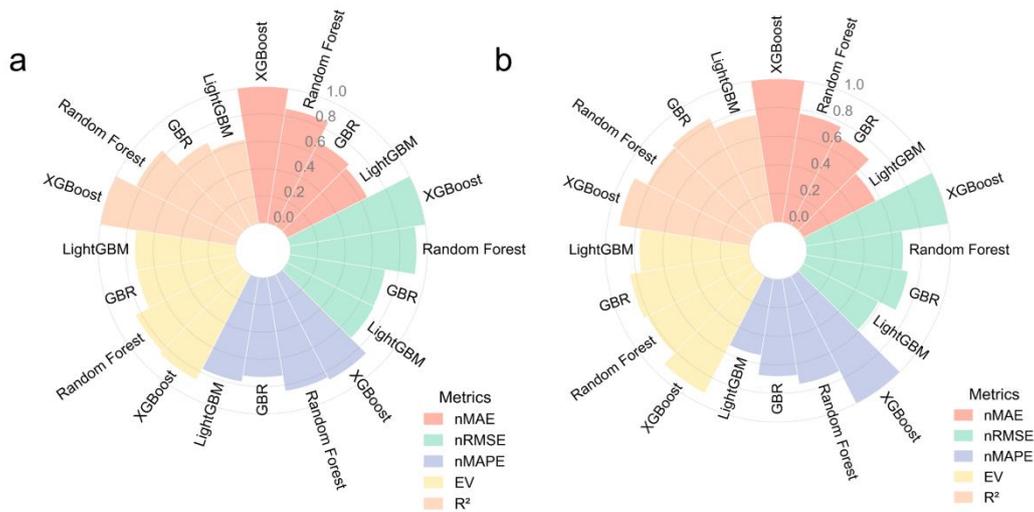

**Fig. 3** Donut bar charts comparing the test performance of five models on (a) Capacity and (b) ICE prediction. All metrics are scaled such that higher values indicate better performance.

## 2.3 Model evaluation on the augmented dataset

To improve predictive performance and overcome the limitations of the initial dataset size, we expanded the dataset to 820 samples using TabPFN-based data augmentation. The augmented dataset was then employed to train four regression models: XGBoost, RF, GBR, and LightGBM. Fig. 4 shows the regression results after augmentation. Fig. 4a and c display the regression predictions versus true values for capacity and ICE using the XGBoost model. The data points are closely clustered along the ideal reference line (y = x), indicating that the predictions closely match the experimental values, demonstrating high accuracy and minimal bias. The fitted regression lines also nearly overlap with the ideal line, suggesting strong generalization ability. Additionally, the histograms above each plot show that both the training and test data distributions are approximately normal, supporting balanced model learning and robust generalization. Fig. 4b and d compare normalized evaluation metrics across the four models. Full performance results are listed in Table 1 (Capacity) and Table 2 (ICE). Compared to results before data augmentation (Tables S3–S4, Supporting Information), all models show improved performance across most metrics. For example, the RMSE of XGBoost decreased from 26.606 to 23.290 in capacity prediction, while the MAE declined from 19.839 to 13.862. In ICE prediction, the $R^2$ of LightGBM increased from 0.762 to 0.849, and the EV of RF improved from 0.820 to 0.846, reflecting enhanced explanatory power and variance capture after data augmentation. We believe these improvements demonstrate that data augmentation effectively increased the diversity and representativeness of the dataset, leading to a more robust model learning process and greater predictive stability.

Additionally, the regression fitting results for the other three models (RF, GBR, and LightGBM) on the augmented dataset are shown in Fig. S3. All these models display visibly improved fitting compared to their performance before augmentation. Specifically, the predicted values are more tightly clustered around the diagonal y = x line, and the fitted regression curves align more closely with the ideal trend, especially for the test data (orange dots). The residual spread is significantly reduced, indicating better predictive stability and less variance. These improvements confirm that the augmented data enables better generalization, allowing the models to more effectively capture the underlying relationships between features and targets.

Although LightGBM slightly outperformed XGBoost on certain ICE-related metrics, XGBoost showed the most consistent and robust performance across both capacity and ICE prediction tasks. Its superior generalization ability and balanced accuracy made it the best choice for further interpretability analysis. Therefore, XGBoost was used for the subsequent SHAP and PDP analyses to clarify the key factors influencing electrochemical performance.

**Table 1** Model evaluation metrics on the test set for Capacity prediction after data augmentation

| ML Model | Model evaluation metrics | | | | |
| --- | --- | --- | --- | --- | --- |
| | Test_MAE | Test_RMSE | Test_EV | Test_$R^2$ | Test_MAPE |
| XGBoost | 13.862 | 23.290 | 0.854 | 0.854 | 4.352 |
| RF | 14.366 | 23.281 | 0.854 | 0.854 | 4.508 |
| GBR | 16.374 | 25.996 | 0.818 | 0.818 | 5.157 |
| LightGBM | 14.329 | 24.264 | 0.842 | 0.842 | 4.500 |

**Table 2** Model evaluation metrics on the test set for ICE prediction after data augmentation

| ML Model | Model evaluation metrics | | | | |
| --- | --- | --- | --- | --- | --- |
| | Test_MAE | Test_RMSE | Test_EV | Test_$R^2$ | Test_MAPE |
| XGBoost | 2.569 | 3.813 | 0.870 | 0.868 | 3.878 |
| RF | 2.720 | 4.121 | 0.848 | 0.846 | 4.098 |
| GBR | 2.623 | 4.087 | 0.854 | 0.849 | 3.928 |
| LightGBM | 2.457 | 4.082 | 0.851 | 0.849 | 3.647 |

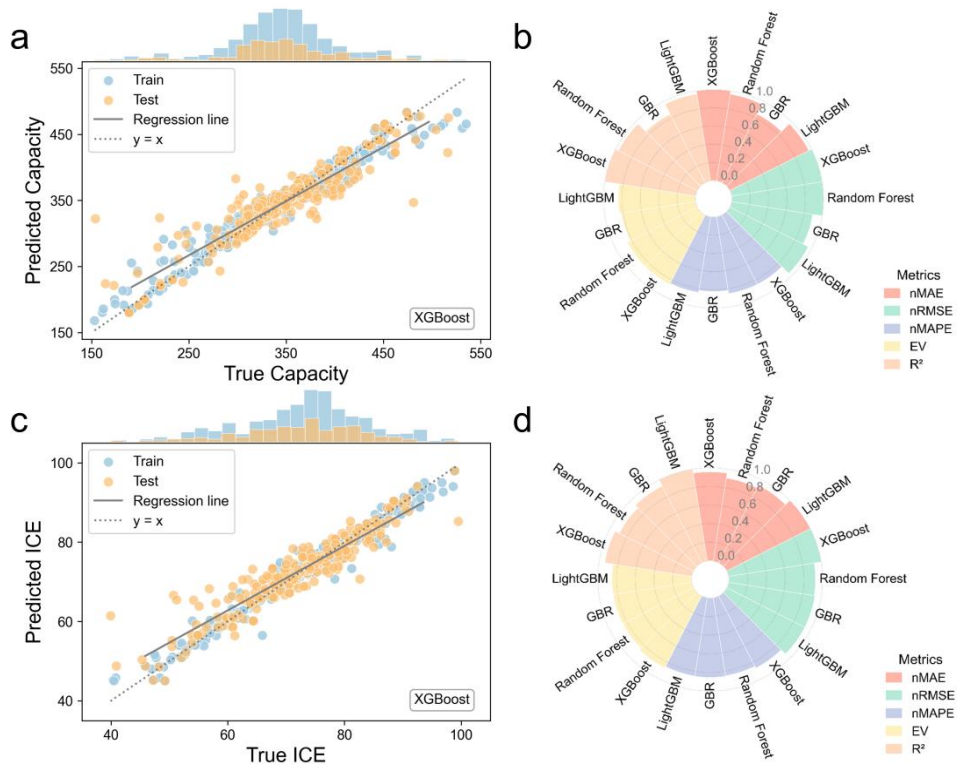

**Fig. 4 Regression fitting and performance comparison of ML models after data augmentation.** Predicted vs. actual values for (a) Capacity and (c) ICE using the XGBoost model, with corresponding data distribution histograms. (b, d) Normalized evaluation metrics for four models across five performance indicators

**2.4 Feature importance analysis**

To explore how individual input features specifically affect the predicted electrochemical performance of hard carbon anodes, SHAP analysis was performed using the XGBoost model. Fig. 5 shows the SHAP summary plots for capacity and ICE. The horizontal bar charts indicate the mean absolute SHAP value of each feature, representing its overall influence on the model output. The nearby rose diagrams display each feature's contribution proportion, helping to determine if the model focuses on a few dominant variables or if the influence is more evenly spread. In the beeswarm plots, each dot corresponds to a single sample's SHAP value for a specific feature; the x-axis shows the positive or negative effect on the prediction, while the color (from red to blue) indicates high or low feature values, illustrating how different feature levels affect the prediction trend.

For capacity prediction, Temperature_2 is the most influential variable, accounting for 28.7% of the total importance, highlighting a dominant role in capacity regulation. Most positive SHAP values are linked to high-temperature samples (red dots), indicating that higher carbonization temperatures improve the predicted capacity. This trend is consistent with previous studies, which show that high-temperature treatment promotes graphitization and structural ordering, facilitating reversible sodium storage and minimizing irreversible trapping [37,38]. Additionally, the SHAP effect of interlayer spacing d002 is mainly influenced by high-value samples (red dots with positive SHAP values), indicating that a larger d002, which means more open layer structures, improves $Na^+$ accessibility and enhances storage performance in certain carbon architectures [39]. Regarding the ID/IG ratio, both high and low values appear across the SHAP axis, while the strongest positive contributions are concentrated in the middle range (purple region), suggesting that moderate structural disorder offers an optimal balance between sodium storage sites and electronic transport pathways [40].

In contrast, ICE prediction shows a more distributed pattern of feature importance. As shown in Fig. 5c, Temperature_2, Time_2, and Temperature_1 are the top three contributors, while electrochemical testing parameters like Current and Initial Voltage also have significant roles. Fig. 5d indicates that higher Temperature_2 and longer Time_2 (red dots) are associated with positive SHAP values, suggesting that more complete carbonization, which reduces oxygen-containing surface groups, helps suppress thick SEI layer formation and thereby improves ICE [41]. On the other hand, higher initial voltage and current are linked to negative SHAP values, probably because they promote side reactions and unstable SEI layers [42]. Additionally, Lc stands out as one of the most influential structural features in ICE prediction. Lower Lc values (blue dots) are linked to positive SHAP contributions, indicating that thinner graphite microcrystallites help reduce irreversible capacity loss by encouraging more uniform SEI formation and limiting deep sodium intercalation [13].

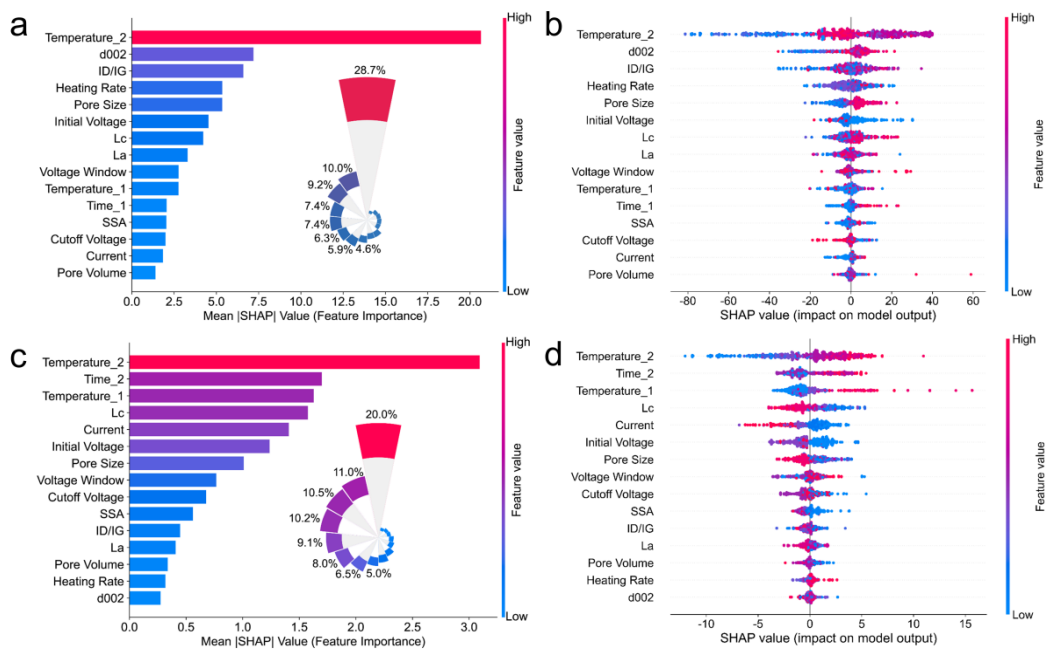

**Fig. 5 SHAP analysis for feature importance using the XGBoost model.** Mean SHAP values showing the global importance of each input feature for Capacity (a) and ICE (c), with embedded rose diagrams illustrating the relative proportion of total feature contributions. SHAP summary plots indicating how individual feature values influence model output for Capacity (b) and ICE (d). Each dot represents one data point; color denotes feature value (red = high, blue = low), and SHAP value on the x-axis indicates direction and magnitude of impact

Fig. 6 shows the PDP for capacity and ICE, providing further insight into the marginal effects and interactions of key variables. By isolating the impact of one feature while averaging over others, PDP analysis helps clarify model behavior and supports understanding the structure-performance relationship. For capacity prediction, Fig. 6a displays a threshold-type nonlinear response with respect to Temperature_2. When Temperature_2 increases from 600 °C to about 1000 °C, capacity remains mostly unchanged; however, there is a sharp rise between 1000 and 1200 °C, followed by a plateau above 1200 °C. This aligns with the SHAP results and suggests that carbonization temperatures beyond a critical point (~1000 °C) promote the development of graphitic domains and increase structural ordering, both of which are beneficial for electron transport and reversible Na+ storage [43]. The PDP of d002 shows a positive correlation, with capacity increasing rapidly in the range of 0.36-0.39 nm, indicating that larger interlayer spacing helps Na$^+$ intercalation by making diffusion easier channels [44]. The ID/IG ratio shows a bell-shaped curve with optimal performance around 1.4–1.8, indicating that moderate structural disorder boosts edge-site density and improves reaction kinetics, while too many defects break the conductive network and weaken the structure integrity [45]. The heating rate exhibits a non-monotonic effect, with performance decreasing at intermediate values and gradually recovering at higher rates, indicating a complex balance between pore formation and structural stabilization. Fig. 6b shows the combined influence of Temperature_2 and the heating rate. Capacity reaches its maximum when Temperature_2 exceeds 1200 °C and the heating rate stays within 4–6 °C min$^{-1}$. Outside this optimal range, capacity decreases sharply, especially at lower temperatures or when heating is too slow, highlighting the importance of coordinating carbonization temperature and heating kinetics for optimal structural development evolution [46].

For ICE (Figs. 6c, d), PDP analysis similarly uncovers nonlinear trends. Both Temperature_2 and Time_2 show critical thresholds for performance improvement, as ICE increases significantly when Temperature_2 exceeds around 1000 °C and Time_2 goes beyond about 2 hours. This highlights the importance of extended thermal treatment in decreasing surface oxygen groups and reducing excessive SEI formation, thus lowering irreversible capacity loss [47]. Current is negatively correlated with ICE, with a steep decline observed above 50 mA g$^{-1}$, indicating intensified side reactions and polarization losses during high-rate testing. The PDP of Lc shows a continuous decrease, with ICE significantly reduced when Lc exceeds approximately 2.5 nm. This suggests that thicker graphite crystallites hinder uniform SEI formation and cause more irreversible sodium trapping due to deeper intercalation sites and uneven surface energetics [48]. Fig. 6d highlights the strong synergy between Time_2 and Temperature_2. ICE remains consistently high only when both parameters reach sufficiently high values. When either the thermal treatment time or temperature is limited, ICE decreases significantly, highlighting that both duration and temperature need to work together to stabilize the electrode-electrolyte interface and prevent parasitic effects reactions [49]. Therefore, we propose that Temperature_2 is the most critical processing parameter, along with Heating Rate and Time_2, which significantly influence the predicted values of capacity and ICE, with a notably higher importance than other features. The subsequent experimental results further confirm the influence patterns of these processing parameters and emphasize their key role in optimizing the electrochemical performance of HC.

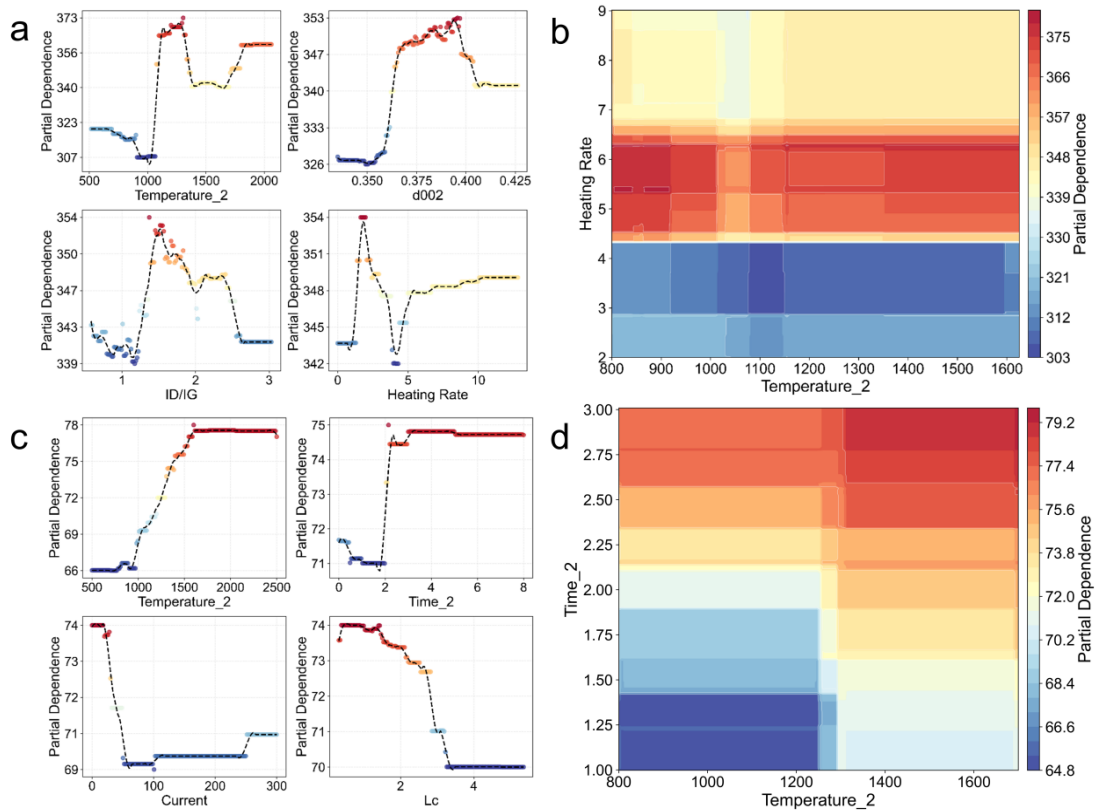

**Fig. 6 PDP analysis revealing the marginal effects and interaction patterns of key features on the predicted Capacity and ICE of hard carbon anodes.** One-dimensional PDP for (a) Capacity and (c) ICE (c), showing how model predictions vary with individual input features (e.g., Temperature_2, d002, ID/IG, Heating Rate). Two-dimensional PDP capturing feature interactions: (b) between Temperature_2 and Heating Rate for (b) Capacity, and (d) between Temperature_2 and Time_2 for ICE. In the one-dimensional plots, the y-axis represents the partial dependence values, and dot colors indicate the magnitude of the model response. In the two-dimensional plots,

the color gradient also reflects predicted values, with red and blue regions corresponding to higher and lower outputs, respectively

**2.5 Experimental validation**

To validate the ML model predictions and explore the dominant influence of key processing parameters on electrochemical performance, four hard carbon anode samples were synthesized from bamboo precursors under controlled conditions. Guided by SHAP and PDP analyses, Temperature_2, Heating Rate, and Time_2 emerged as the most influential factors, with Temperature_2 being the most critical parameter affecting both capacity and ICE. Accordingly, the experimental design focused on these three parameters while maintaining other conditions constant. As detailed in Table S5, the synthesis parameters were strategically designed with HC-1 as the baseline condition, and HC-2, HC-3, and HC-4 systematically varied Heating Rate (2 °C min$^{-1}$ vs 5 °C min$^{-1}$), Temperature_2 (1000 °C vs 1300 °C), and Time_2 (1 h vs 2.75 h), respectively, to validate the ML-predicted parameter effects.

The structural features of the four hard carbon samples were analyzed using Raman spectroscopy and SEM to understand how synthesis conditions relate to material properties. Fig. S4 displays the Raman spectra of all samples, showing the characteristic D band (~1350 cm$^{-1}$) and G band (~1590 cm$^{-1}$), which are linked to disordered and graphitic carbon structures, respectively. The calculated ID/IG ratios are 1.795, 1.810, 2.206, and 1.846 for HC-1, HC-2, HC-3, and HC-4, respectively. HC-3, made at the lowest carbonization temperature (1000 °C), has the highest ID/IG ratio (2.206), indicating more structural disorder. This exceeds the optimal range (1.4~1.8) predicted by PDP analysis and relates to its lower electrochemical activity observed later. In contrast, HC-1, HC-2, and HC-4, produced at a higher temperature (1300 °C), show lower ID/IG ratios, closer to or within the ML-predicted optimal range, indicating improved graphitization that supports better electronic conductivity and sodium storage performance [50]. SEM images (Fig. S5) show that all samples have typical hard carbon structures with irregular, broken particle shapes and sizes ranging from a few to tens of micrometers. The particles have rough surfaces with noticeable porosity, which is typical of biomass-derived hard carbons after high-temperature carbonization [51]. While the external morphologies seem similar across all samples, the different synthesis parameters mainly influence the internal microstructure, especially the level of structural disorder and graphitization, as shown by the distinct ID/IG ratios observed in the Raman analysis. The structural characterization confirms the ML model's focus on carbonization temperature as the main factor, where the Temperature_2 difference between HC-3 (1000 °C) and the other samples (1300 °C) leads to clearly different levels of structural disorder that directly impact the subsequent electrochemical performance.

Fig. 7a shows the initial galvanostatic charge-discharge profiles of the four hard carbon samples tested at 50 mA $g^{-1}$ between 0.001-3.0 V vs. $Na^+$/Na, where all samples exhibit typical hard carbon characteristics with a sloping region above 0.2 V and a low-voltage plateau below 0.2 V, corresponding to $Na^+$ insertion into defects/micropores and intercalation between graphene layers, respectively [52]. The initial discharge capacities follow the order HC-1 (324.61 mAh $g^{-1}$) > HC-2 (304.63 mAh $g^{-1}$) > HC-3 (284.92 mAh $g^{-1}$) > HC-4 (279.62 mAh $g^{-1}$), with corresponding ICE of 80.0%, 78.7%, 78.4%, and 79.2%, respectively (detailed performance data are provided in Table S6). The rate capability tests (Fig. 7c) show that HC-1 maintains the highest specific capacity across all current densities from 50 to 500 mA $g^{-1}$, while all samples recover capacity well when returned to 50 mA $g^{-1}$, indicating excellent structural stability during high-rate cycling. Short-term cycling performance (Fig. 7b) displays the initial behavior of all samples, while long-term cycling results over 300 cycles (Fig. 7d) reveal different capacity retention behaviors, with HC-2 showing the best stability (89.45% retention), followed by HC-3 (89.31%), HC-4 (88.58%), and HC-1 (86.14%). These experimental results strongly confirm the ML model predictions: HC-1, synthesized with optimal high-temperature conditions (1300 °C, 5 °C $min^{-1}$, 2.75 h), achieves both the highest capacity and ICE as predicted, while HC-3 prepared at a lower temperature (1000 °C) exhibits significantly reduced capacity (284.92 vs 324.61 mAh $g^{-1}$) and lower ICE (78.39% vs 79.99%), confirming the critical role of carbonization temperature above 1000 °C as identified by PDP analysis. The ranking based on capacity and ICE closely matches the synthesis parameter optimization trends identified by SHAP analysis, where Temperature_2 dominates performance, demonstrating the reliability and practical application of the ML approach for guiding experimental design and parameter optimization in hard carbon anode development.

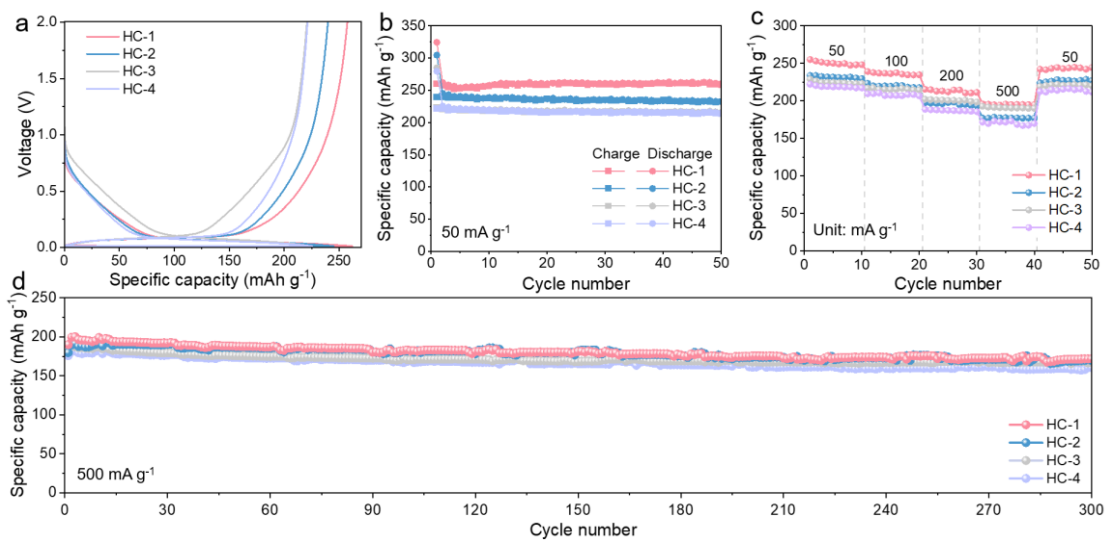

**Fig. 7 Electrochemical performance.** (a) Galvanostatic charge-discharge, (b) cycling performance, (c) rate, and (d) Long-term cycling performance of HC-1, HC-2, HC-3, and HC-4

## 3. Conclusion

In this work, four ML models (XGBoost, RF, GBR, and LightGBM) were used to predict the capacity and ICE of hard carbon anodes for SIBs, with robustness under small-sample conditions improved through TabPFN-based data augmentation. Feature importance analysis of the top-performing model showed how key features influence electrochemical performance. Among the models, XGBoost achieved the highest accuracy, with $R^2$ values of 0.854 and 0.868 and RMSE values of 23.290 mAh $g^{-1}$ and 3.813% for capacity and ICE, respectively. SHAP and PDP analyses identified Temperature_2 as the key feature affecting both metrics, while also clarifying the roles of other factors. Based on these insights, four hard carbon samples were synthesized from bamboo precursors with optimized parameters and underwent structural and electrochemical testing. The close match between predicted and experimental results highlights the reliability and usefulness of the ML framework. However, current hard carbon datasets remain limited, especially regarding microstructural and interfacial descriptors such as detailed pore-size distributions and defect types, which restricts the completeness of feature importance analysis. We also plan to address these limitations in future work to develop a more accurate framework for creating high-performance hard carbon anodes.

## Data availability

Data will be made available on request.

## Declaration of competing interest

The authors declare no competing interests in this work.

## Acknowledgements

We thank the financial support from Guangxi Science and Technology Program (2024AB08156), Natural Science Foundation of Shanghai (25ZR1401102), and Science and Technology Program of Guangzhou, China (SL2024A03J00326).